\documentclass[preprint,showpacs,prb,nobibnotes]{revtex4} 

\usepackage{graphicx}

\begin{document}

\title{Binding Energies in Benzene Dimers: Nonlocal Density Functional Calculations}
\author{Aaron Puzder}\altaffiliation[Present address:\ ]{Lawrence Livermore
National Laboratory, Livermore, California 94550} 
\author{Maxime Dion}
\author{David C. Langreth}
\affiliation{Center for Materials Theory, Department of Physics and Astronomy\\
Rutgers University, Piscataway, New Jersey 08854-8019}
\date{\today}

\begin{abstract}
The interaction energy and minimum energy structure for different 
geometries of the benzene dimer has been calculated using 
the recently developed nonlocal correlation energy 
functional for calculating dispersion interactions. 
The comparison of this straightforward and relatively 
quick density functional based method with recent calculations 
can elucidate how the former, quicker method might be exploited in 
larger more complicated biological, organic, aromatic,
and even infinite systems such as molecules physisorbed on surfaces,
and van der Waals crystals.
\end{abstract}


\maketitle

\section{Introduction}
Nonempirical density functionals of semilocal or generalized gradient 
approximation (GGA) type have met with a good degree of success when 
applied to either isolated molecules\cite{GGAmolecule} or dense solid-state 
systems.\cite{GGAsolid}   Generally such approximations do not include the 
long range dispersion or van der Waals (vdW) interaction, and hence fail 
or at best give sporadic results for sparse matter or for the wide range 
of molecular complexes where the dispersion interaction is an important 
component of the binding.  Even semiempirical gradient functionals typically 
fail for such systems, in the sense that their parameters must be varied 
from system to system, and thus lose most of their predictive power.

Over the course of the last several years the Chalmers-Rutgers 
collaboration\cite{rydbergPRL,langrethIJQC,dion} has developed a nonemperical 
density functional for the correlation energy which showed considerable 
promise in alleviating the above undesirable situation.  This correlation 
energy functional is {\it not} of the GGA type, but involves a fully 
non-local integration over pairs of densities at distant points.  Nevertheless 
it is very quick to evaluate and does not significantly lengthen a simple 
GGA density functional calculation for a sufficiently large system.  As 
proposed\cite{dion}, this nonempirical correlation functional has no 
corresponding nonempirical exchange functional. Because the 
Zhang-Yang\cite{revPBE} GGA ``revPBE'' exchange functional had been fitted 
to exact exchange calculations and because we indeed found that it gave 
the best representation of exchange for this type of problem of all the 
GGA exchange functionals tried, we adopted it as part of our recommended 
procedure.\cite{rydbergPRL,dion} This total van der Waals density functional 
(vdW-DF) has been applied to layered structures where it produced reasonable 
agreement with experimental results.  Such layered systems, namely bulk 
graphite, molybdenum sulfide, boron nitride\cite{rydbergPRL} as well as 
the corresponding layered dimers\cite{langrethIJQC} are all ones in which 
GGA fails completely.  Going to a more general geometry,\cite{dion} tests 
were made on rare-gas dimers as well as the benzene dimer in the 
particularly simple ``sandwich'' or ``atop-parallel'' geometry.  

The benzene dimer represents an ideal testing ground  for 
new correlation density functionals because of the wealth 
of wave-function calculations\cite{hobza1,hobza2,jaffe,hobza3,tsuzuki1,sinnokrot1,sinnokrot2} 
on this system in different geometries.  These represent both M\o ller-Plesset 
theory (MP2) and coupled-cluster theory (CCSD(T)); the more recent of the 
latter represent the current state-of-the-art and are extremely demanding 
computationally.  Their expected accuracy has been well discussed by their 
practitioners.  These theories based on wavefunction calculations may be 
adopted as giving reference values, in lieu of conclusions from experimental 
work on such weakly binding systems, which require a number of assumptions 
to deduce binding structures and energies.\cite{bornsen,grover,krause,arunan,felker} 
More recently there have been a rather full set of MP2 and CCSD(T) 
calculations\cite{sinnokrot3} on the benzene-phenol, -toluene, -fluorobenzene, 
and -benzonitrile complexes.  This has provided an opportunity for further 
testing of the functional on a wider variety of well referenced systems.  
Such a study has recently been made\cite{timosub} which 
shows promise for the functional for different geometries of those four 
systems as well.

In this paper, we make a thorough study of the behavior of this van der
Waals density functional\cite{dion} (vdW-DF) for a number of different benzene
geometries.  We calculate the interaction energy and lowest energy 
structure for different geometries of the benzene dimer  with a converged 
calculation using the nonlocal correlation energy functional within the 
framework of a norm-conserving pseudopotential plane wave code.  The 
comparison of this straightforward and relatively quick density functional 
based method with other methods can determine potential errors in this 
method and how such a method might be utilized in larger, more complicated 
biological, organic, and/or aromatic systems.  We address the exchange 
functional   by studying the effect of replacing the GGA exchange functional 
with a full Hartree-Fock (HF) calculation.  Indeed, we find that slightly 
weaker repulsion of the full HF calculation gives improved bonding distances 
along with stronger binding.  The various tradoffs between the two methods 
are studied and discussed.  We benchmark this functional on the benzene 
dimer system where a good amount of data is available from other calculations
and where generated conclusions still conflict with one another.  By comparing with these systems 
in which state-of-the-art results exist, we demonstrate that we can utilize 
this functional for much larger biological and organic systems which go beyond 
wavefunction calculations.

\section{Computational Methods}
Our calculations were performed using density 
functional theory (DFT) with nonlocal correlation energy. 
Our DFT calculations follow the previous prescription
(which we continue to abbreviate as vdW-DF) 
for calculating binding energies in van der Waals bonded systems
with this particular nonlocal correlation functional\cite{dion} 
\begin{equation}
E[\rho] = T_\mathrm{s}[\rho] + V_\mathrm{pp}[\rho]+J[\rho] + E_\mathrm{x}[\rho] +
 E_\mathrm{c}^\mathrm{L}[\rho] + E_\mathrm{c}^\mathrm{NL}[\rho]
\label{vdwdf}
\end{equation}
where E is the total energy functional of the dimer or monomer, 
$T_\mathrm{s}$ is the single-particle kinetic-energy functional, $V_\mathrm{pp}$ is the 
ionic pseudopotential functional, $J$ is the Coulombic interaction
functional, $E_\mathrm{x}$ is the revPBE flavor\cite{revPBE} of the generalized 
gradient approximation (GGA) exchange functional, 
$E_\mathrm{c}^\mathrm{L}$ is the local contribution to the 
total correlation energy, and $E_\mathrm{c}^\mathrm{NL}$ is the nonlocal contribution to 
the total correlation energy, 
\begin{equation}
E_\mathrm{c}^\mathrm{NL} = 
\frac{1}{2} \int\!d^3r_1\,d^3r_2\, \rho(\vec r_1) 
\phi(\vec r_1, \vec r_2)\rho(\vec r_2).
\label{twopoint}
\end{equation}
The kernel can be written in a form allowing rapid evaluation
\begin{equation}
\phi(\vec r_1, \vec r_2)=   \tilde\phi(Rf(\vec r_1),Rf(\vec r_2)),
\end{equation}
where $f(\vec r_i)$ is a function only of $\rho(\vec r_i)$
and $|\nabla \rho(\vec r_i)|$ for $i=1,2$, and $R=|\vec r_1 - \vec r_2|$.
The details are given in Ref.~\onlinecite{dion}, which introduces
a sum and difference variable decomposition that provides still further
simplification.

Although similar in spirit to the 
calculations describing the nonlocal correlation functional in a previous
work\cite{dion}, we utilize 
some notable exceptions.  First, we use norm-conserving pseudopotentials 
of the Troullier-Martins type\cite{troullier} for $V_\mathrm{pp}$ within the framework of 
the {\it abinit} code.\cite{abinit}  Second, we use the 
Perdew-Burke-Ernzerhof (PBE) exchange-correlation potential\cite{pbe} 
as well as charge 
densities derived from that calculation, {\it i.e.}\ $\rho$ is $\rho_\mathrm{PBE}$ for 
the calculation of each functional.  We calculated a number of representative
binding energies using other DFT charge densities, but found negligible changes
in all results.  We calculate $E[\rho]$ using revPBE, then subtract the GGA 
correlation energy before adding 
$E_\mathrm{c}^\mathrm{L}$ with a local density approximation (LDA) correlation 
functional.  Whereas each term is calculated using {\it abinit}, 
we calculate the nonlocal correlation energy 
$E_\mathrm{c}^\mathrm{NL}$ as a post process 
calculation to obtain our vdW-DF solution. 

Except for the evaluation 
of $E_\mathrm{c}^\mathrm{NL}$, we employ a plane\-wave 
approach with periodic boundary conditions applied to a supercell with 
large enough spacings such that no spurious interactions exist between 
periodic replica.  All dimers are placed in a box with 15 to 20~\AA 
 ~of vacuum between the clusters.  This amount of vacuum proved sufficient 
for the total charge density to approach zero ($10^{-9}$) well 
before the the supercell edge.  The kinetic energy cutoff used is at 
least 50~Ry corresponding to about $2\times 10^5$ plane waves in a cell of 
21~\AA ~per side.  We found that larger supercells and higher energy cutoffs had 
negligible effect (less than $0.05\%$) on the total binding energy
of any structure.  The post process calculation of $E^\mathrm{NL}_c$ was evaluated
on a real space grid at a size equal to the supercell and with grid spacings equal to the 
Fourier transform spacings in the planewave calculations.  $E^\mathrm{NL}_c$ was 
evaluated at each grid point and numerically integrated in a manner described in 
greater detail in a previous work.\cite{dion} Although $E^\mathrm{NL}_c$ is not 
calculated self-consistently at this time, other DFT charge densities yield no
change in the results, suggesting self-consistency would not change any result
appreciably or overall conclusions.

\begin{figure*}[t]
\begin{center}{\includegraphics[width=.96\textwidth]{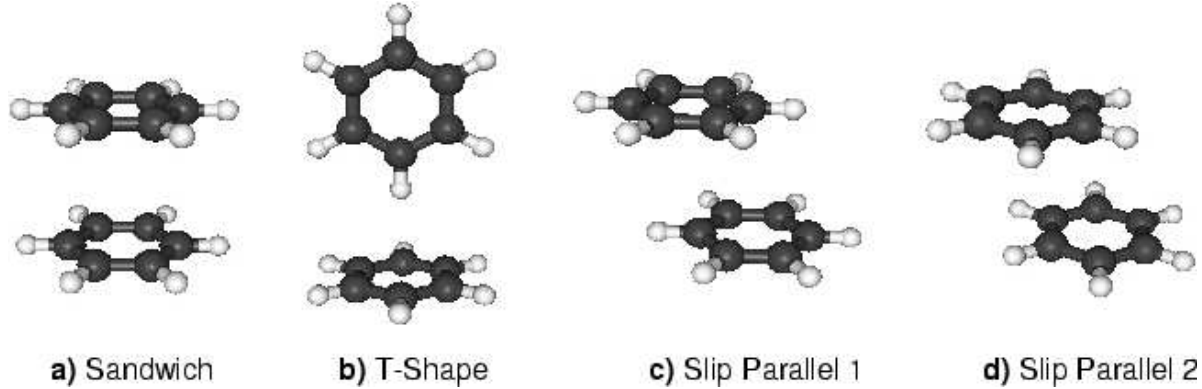}}\end{center}
\caption{The benzene dimer in various geometries including a) the on top parallel (sandwich) structure b) the T-shape, orthogonal structure c) the slip parallel structure in which one benzene molecule is slipped along the carbon-hydrogen bond and d) the slip parallel structure in which one benzene molecule is slipped perpendicular to a carbon-carbon bond.  Carbon is black and hydrogen is white in the ball and stick representation. }
\label{fig1}
\end{figure*}

We calculated the binding energies of two distinct forms of the benzene 
dimer, the parallel and T-shape dimers (see Fig.~1).  For parallel geometries, 
we considered the on-top parallel sandwich structure (Fig.~1(a)), and 
two different slip parallel structures (Fig.~1(c) and (d)).  We calculated 
the total binding energies for the slip parallel structures by starting 
from the graphite-like AB stacked (minimum energy) geometry and moving one 
monomer vertically until a 
local minimum was reached.  We subsequently moved that monomer horizontally 
along a C-H bond once that vertical minimum point had been ascertained.  
We then moved that same monomer in each direction to verify that this
was a local minimum.  Once 
we found this minimum (Fig.~1(c)), we started from a new slip parallel 
structure defined by slipping towards the C-C bond.  We moved vertically 
and horizontally until a new slip parallel minimum was found (Fig. 1(d)).~ 
By slipping in two orthogonal directions, we hope to find the lowest 
possible slip parallel structure.  Every benzene molecule was locked at a
fixed structure of 1.397~\AA ~C-C length and 1.079~\AA ~C-H length.  
Changing these structures, for example, to the minimum LDA or GGA 
geometries made no difference ($< 0.0005$~kcal/mol) to any final binding 
energy results.

\section{Results}
Fig.~2 and Fig.~3 show the DFT, nonlocal correlation (vdW-DF) results of three structures 
including the parallel on top (sandwich) structure and two other potential lowest energy 
structures, the T-shape and the slip parallel structures.  Each of these results 
is compared with recent CCSD(T) and MP2 calculations.\cite{tsuzuki1}  
The sandwich structure dimer is similar with a recent calculation using 
this same nonlocal correlation functional\cite{dion} in that the minimum energy point
and the value of the binding energy at that point agree, but
disagrees slightly at larger separations of the dimer. 
The results here used very large box sizes and very refined
integration cutoff parameters.  The results for benzene dimers in this work should 
be taken as the results generated from this methodology.
\begin{figure*}[t]
\vspace*{1em}
\begin{center}{\includegraphics[width=.96\textwidth]{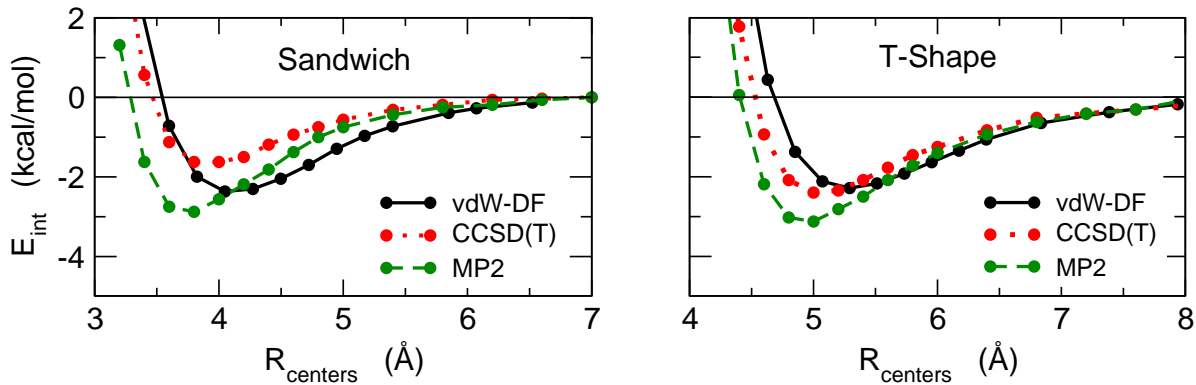}}\end{center}
\caption{Interaction energies ($E_\mathrm{int}$)
of  the indicated benzene dimers using vdW-DF (Eq.~\ref{vdwdf}),
 compared with previous M{\o}ller-Plesset perturbation (MP2)
  and coupled-cluster (CCSD(T)) calculations.\cite{tsuzuki1}
 These DFT calculations used  nonlocal correlation energies\cite{dion} and
 revPBE exchange energies.\cite{revPBE}
 The abscissas ($R_\mathrm{centers}$) give
  the center-to-center distance
 between the benzene monomers. 
 }
\label{fig2}
\end{figure*}

\begin{figure}[!t]
\vspace*{1em}
{\includegraphics[width=.42\textwidth]{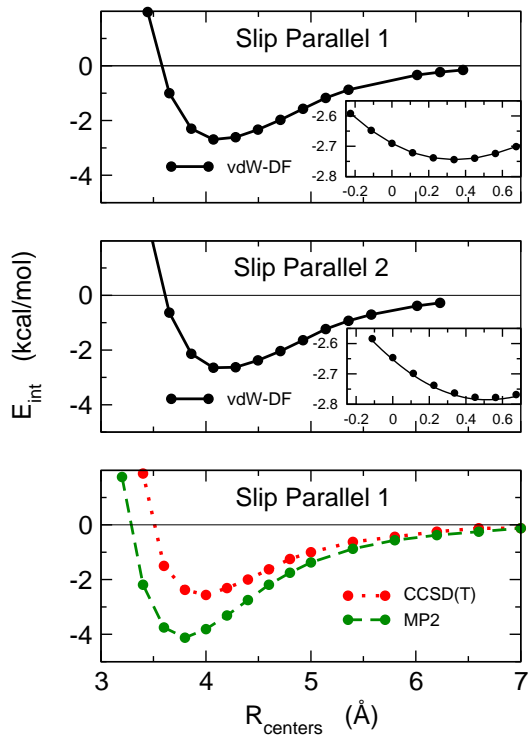}}
\vspace*{8em}  
\caption{vdW-DF interaction energies of benzene dimers in the indicated
configurations.
The larger graph in each of these two panels shows the effect of vertically
(see Fig.~1) separating of the monomer planes at a fixed horizontal slip
distance of 1.398 {\AA} 
between the monomer axes. The insets show the effect of
varying this slip distance away from the energy minimum found from the
previous movement: the total horizontal 
slip distance is 1.398 {\AA} \textit{plus} the
abscissa of the respective inset.  The abscissas of first two larger graphs are the
values of the monomer intercenter distance 
$R_\mathrm{centers}$ \textit{before} the horizontal displacement
indicated in the insets.
The third panel shows for comparison  MP2 and CCSD(T)
calculations,\cite{tsuzuki1} where the vertical and horizontal distances were
increased simultaneously in a fixed ratio (see text).
}
\label{fig3}
\end{figure}

In each case, we find that our calculations predict a significantly 
larger dimer separation, up to $18\%$ larger when compared with CCSD(T) and MP2.
We find that our binding energies agree extremely well with the CCSD(T) calculation for 
the magnitude of these binding energies, both of which are significantly 
less than those generated from the MP2 calculation.  Additionally, we agree 
with these previous calculations as well as more recent MP2 and CCSD(T) 
calculations\cite{sinnokrot1,sinnokrot2} that, despite being nearly isoenergetic, 
the slip parallel is the lowest energy structure.\cite{tsuzuki1}

Although the dimer separation ($R_\mathrm{centers}$) is always given here as the 
distance from the center of the benzene rings regardless of the structure, in the 
slip parallel case the previous CCSD(T) and MP2 results, with which we compare,
were generated by 
fixing the angle between the center of mass line and the horizontal at 
$63^{\circ}$.\cite{tsuzuki1}  In our calculation, we moved one benzene 
ring vertically and 
and then horizontally as in Fig.~1(c)  until we obtained a minimum.  
Additionally, we moved one benzene ring 
at an angle of 30$^{\circ}$
to this movement, but in the same
plane (Fig.~1(d)). 
Surprisingly, 
we found a significantly stronger binding energy when moving in this 
alternate direction, suggesting that the structure represented by Fig.~1(d) 
is the lowest energy structure for the slip parallel geometry.  This result is especially
noteworthy as it represents a difference in stacking between the lowest energy 
structure of graphite and of the benzene dimer.  Although 
disagreeing with some previous assumed slip parallel structures,\cite{jaffe,hobza3} 
this geometry is actually similar to other previous studies\cite{hobza3} including 
the structure found by recent and 
more complete CCSD(T) calculations\cite{sinnokrot1,sinnokrot2} although no discussion or 
comparison was made with the alternate structure in these works.

One of the major benefits of using the nonlocal correlation functional 
for organic, aromatic, and $\pi$-bonded systems is its compatibility with 
standard DFT planewave codes and the relative speed and lack of computer 
power needed.  Indeed, these benzene dimers pushed CCSD(T) nearly to its peak. 
Even the less computationally demanding
MP2-R12/A calculations were reported to have required
two weeks on four processors.\cite{sinnokrot1}  Calculations of the
later type are important, not only in their own right, but are also
also often used as input for a clever procedure to extract a best estimate of
the basis set limit for the even more demanding CCSD(T) calculations.
In contrast
the the evaluation of $E_\mathrm{c}^\mathrm{NL}$ done here took 
a time\cite{optimize} of the same order as
corresponding GGA calculation
that proceeded it (about an hour per point on a
Dual Athalon MP 2000+ processor), and the former will take comparably 
less time as the system size increases.
Therefore, the benzene dimer represents a starting point for such 
systems within the present implementation, rather than nearly a limiting case 
as in CCSD(T).   Although another DFT based methodology using 
symmetry-adapted perturbation 
theory\cite{misquitta,szal1} is reported\cite{szal2}  
to have been recently
optimized such as to
offer comparable speeds for the benzene dimer, 
it nevertheless scales as the fifth power\cite{szal2}
of the basis size, and hence in contradistinction to the present DFT
scheme, will rapidly become unmanageable as the system size increases. 
Other coupled-cluster schemes such as those placing effort into linear 
scaling might also offer
comparable speeds, but that such a scheme would generate consistent results
in van der Waals systems has not yet been indicated. Therefore, 
this current methodology
offers the only  method we know that gives a calculation speed
comparable to that of an ordinary GGA calculation as the system size
increases, and in order to 
utilize it fully, understanding the  small disagreement of the 
predicted dimer separation in this work with the previous CCSD(T) calculations 
is vital to increase the applicability of this method.

\section{Discussion}
The results of the vdW-DF calculation when compared with the 
CCSD(T) and MP2 results are quite illuminating.  First we note that 
the binding energy values are extremely consistent with the CCSD(T) results, 
much closer in the two minimum energy structures than the MP2 calculations.  
Furthermore, both these results are consistent with the CCSD(T) results in 
that both show that the T-shape and the slip parallel structures to be nearly 
isoenergetic with the slip parallel slightly lower.  This result is 
a strong indicator that this method might be the most efficient for 
calculating binding energies in van der Waals systems.

CCSD(T) is considered the best and most complete answer to date but suffers 
from some serious scaling constraints as the system size increases.
One can obtain useful results by using the best basis set possible,
and then applying a clever method for estimating the basis set limit
by making use of an MP2 calculation which can be made with a much larger
basis set.  The results used in our plots here\cite{tsuzuki1}
used this method. A more recent calculation\cite{sinnokrot2}
using this method with even
a better basis set (aug-cc-pVQZ$^*$) and using an MP2 calculation essentially
at the basis set limit\cite{sinnokrot1} yields qualitatively similar results 
varying by a few tenths of a kcal/mol.
  Thus when searching for alternatives, 
it is clear that our method gives closer energies to CCSD(T) compared 
with MP2 at the basis set limit, while yielding inconsistent 
separations.  Unfortunately, in addition to these slightly larger separations, this 
method also gives an inconsistent result in that the sandwich structure 
is lower in energy than the T-shape structure. To understand why 
requires greater analysis.

In addition to the interplay of all the typical interactions,
the binding energy of benzene dimers is influenced by two competing 
interactions not necessarily as prominent as in other physical systems: 
the attractive dispersion interaction and the repulsive 
electrostatic quadrupole-quadrupole interaction.  How any method deals with these two 
interactions, tiny when compared with their total energies, is thus vital in 
determining which structure it might favor.  Parallel 
structures, both the sandwich and the slip parallel, feature a more 
dominant quadrupole-quadrupole repulsion interaction than the T-shape 
structure, an interaction that tends to decrease  
the total binding energy when dispersion is not fully
taken into account.  If the counterbalancing dispersion interactions are 
neglected or only partially included, T-shape structures 
may then be found as the more stable structure.  Conversely, any 
method that underestimates quadrupole-quadrupole interactions such as DFT based methods, or 
any method that overestimates nonlocal correlation effects, would bias the result towards 
favoring parallel structures.  

The MP2 perturbation method is a sensible method
for dealing with these two effects in a controllable manner.  The correlation energy
calculated through the MP2 method counterbalances the quadrupole-quadrupole interaction. 
The completeness of the Gaussian basis set used and the inclusion 
of disperse polarizability functions tends to increase the magnitude of 
the total electron correlation energy in MP2 calculations and thus the 
attractive dispersion interactions.  As the basis sets are less complete, 
the dispersion energy will not compensate the electrostatic effect fully 
and T-shape structures will be found to be more favorable.  Methods based
on multipole expansion\cite{pawliszyn,price} and underconverged MP2 
methods\cite{hobza1,hobza3} led to this exact trend 
in the 1980s and the early to mid 1990s.  
As the capability of performing calculations with a more converged basis 
set occurred, the planar geometries felt the effect to a greater degree, 
while the T-shape geometries changed in energy negligibly.  Eventually, 
the capability to do a converged basis set HF and MP2 calculation
led to such an increase in binding energy 
(more negative interaction energy) for the planar 
structure, that planar geometries were recognized as the more favorable 
energy structure.\cite{jaffe,tsuzuki1,sinnokrot1,sinnokrot2}

This historical analysis demonstrates how different methods will bias 
which nearly isoenergetic system will be found to be the lowest energy 
structure.  In our case, we note that because the slip parallel dimer was 
found as the lowest energy structure, our method for including nonlocal 
correlation energy is consistent with the converged MP2 
and the CCSD(T) calculations, \textit{i.e.} our nonlocal correlation functional 
captures an amount of correlation consistent with converged MP2 calculations.  
However, because we perform a DFT calculation 
and use DFT densities, the quadrupole moment is 
underestimated by $\approx 18$ to $23\%$,\cite{meijer} giving a
quadrupole repulsion in the sandwich configuration
that is substantially too small.  Thus our sandwich binding energies 
will be too large because not enough repulsion is considered,
while the T-Shape is largely unaffected because of the larger distances.
Because our T-shape binding energy predictions were less than $0.09$~kcal/mol weaker 
 than the sandwich binding energy predictions, this may be one source of the 
physical discrepancy of the sandwich binding energy being found stronger than the 
T-shape binding energy.  Furthermore, this result also demonstrates a caveat for using this method 
when the principal interaction is electrostatic and not van der Waals.
Possibly this defect could be corrected \textit{ad hoc} by applying a density
correction method\cite{baerends94,tozer} as done in Refs.~\onlinecite{misquitta}
and \onlinecite{szal1}, but we have not yet explored this possibility.  
Any method that features a complete quadrupole-quadrupole 
interaction energy and dispersion energy may reveal that the T-shape is the more energetically 
favorable than the slip parallel, but clearly both distinct structures are 
local minimums and should thus both occur in nature.

One trend that we observe is the consistent overestimation of dimer 
separation in benzene compared with CCSD(T).  
The possibility of this
particular result being a specific trait of benzene 
and the use of the revPBE exchange energy are investigated as two 
potential explanations for this discrepancy.

With the lack of conclusive experimental equilibrium distances in gas 
phase benzene, we performed a similar comparison with a system that has 
a known experimental dimer separation.  This comparison may also reveal 
if our overestimation of dimer separation is specific to benzene or a 
general trait of this methodology.  To this end, we calculated the geometry 
and binding energy of the argon dimer.  Similar to the previous published 
results\cite{dion}, we obtain a binding energy of 0.45~kcal/mol at 3.97~\AA~for 
the argon dimer.  This equilibrium distance overestimates the accepted 
experimental dimer separation by $\approx$~0.2~\AA ~or $6\%$.  Unfortunately, 
this result is inconclusive in explaining our discrepancy in benzene as it 
still overestimates the accepted argon experimental value, but not nearly 
as much as the assumed overestimation in benzene.  

\begin{figure}[t]
\vspace*{1em}
\begin{center}{\includegraphics[width=.48\textwidth]{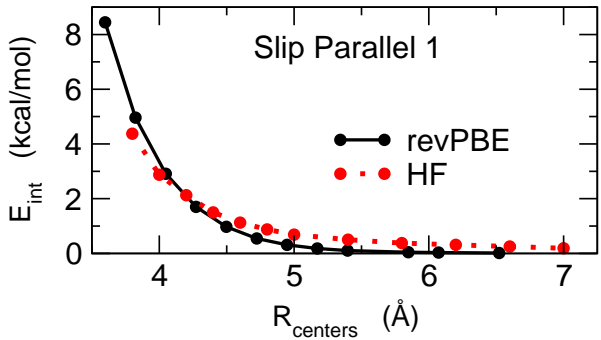}}\end{center}
\caption{Binding energies of the sandwich structure benzene dimer using two methods when correlation is not included, the Hartree-Fock (HF) method and the revPBE method with exchange only and no correlation.}
\label{fig4}
\end{figure}

Thus, we analyzed the effect of using revPBE exchange energies in our 
binding energy calculation.  Fig.~4 shows revPBE exchange binding energies 
compared with a Hartree-Fock calculation for the sandwich structure.\cite{tsuzuki1} 
For the sandwich structure, we see that revPBE is steeper than HF, and 
approaches zero more rapidly.  These differences suggest that the revPBE 
energy, \textit{not} the nonlocal correlation energy correction is responsible 
for the discrepancy in dimer separation with MP2 and CCSD(T).  Furthermore, 
the revPBE exchange results also appear to be responsible for the slow 
approach towards zero from below in the total binding energy.  Fig.~5 
and Fig.~6 show the effect of using Hartree Fock as opposed to revPBE exchange
on the total binding energy.  In these results, the 
correlation energy is not changed at all, simply the revPBE exchange-only 
energy is subtracted from the total energy and HF energy is added.  Both 
the dimer separation and the slow approach towards zero problems simultaneously 
disappear for the sandwich structure: the minimum displacement shifts 
inward by nearly 0.43~\AA ~while the binding energy approaches zero from 
below more quickly and more consistently with the MP2 and CCSD(T) result.  
Also disappearing is the rather large energy difference between the two 
slip parallel equilibria, with the slip-parallel-2 geometry only 0.005 
kcal/mol lower (\textit{i.e.}, more stable).  Unfortunately, the change 
in exchange energy tends to increase the magnitude of the binding energy, 
placing it further away from the CCSD(T) calculations, although still 
comparable to the MP2 calculations.  Based on this result, the very slight 
differences in revPBE with Hartree-Fock (HF) seem to be the root of most 
discrepancies, especially dimer separation, although revPBE is the most 
consistent exchange density functional of those tested\cite{dion} for 
representing exchange-only effects.  This  difference can slightly alter 
equilibrium geometries in benzene dimers, and thus perhaps HF is a better 
starting point.  We have redone our exchange only calculations for each benzene dimer 
with HF calculations instead of revPBE.  In every case, we obtain a minimum 
point consistent with CCSD(T), and energies consistent with MP2.  Additionally, 
as HF and revPBE differ in slightly different manners in each system, we find 
that the T-shape dimer now binds much more strongly than the sandwich dimer.  
\begin{table}
\vspace*{2em}
\begin{center}\begin{tabular}{lcccccc}
\hline
			&Sandwich	&T-Shaped	&Slip-Parallel-2	&Angle	\\ \hline
vdW-DF			&4.1		&5.3		&4.4		&64$^{\circ}$	\\
vdW-DF(HF)		&3.8		&4.9	  	&3.7		&64$^{\circ}$	\\
MP2$^\mathrm{a}$	&3.7		&4.9		&3.8		&65$^{\circ}$	\\
CCSD(T)$^\mathrm{b}$ 	&4.1		&5.0		&4.0		&63$^{\circ}$	\\
Experiment$^\mathrm{c}$	&		&4.96		&		&		\\ \hline
\end{tabular}
\caption{Equilibrium center-to-center distance ($R_\mathrm{centers}$)
 in \AA\ (and angle to basal planes for slip-parallel
configuration) obtained by various methods: $^\mathrm{a}$Ref.~\onlinecite{sinnokrot1};
$^\mathrm{b}$Ref.~\onlinecite{hobza3};	$^\mathrm{c}$Ref.~\onlinecite{arunan}.
\label{distances}
}
\end{center}\end{table}
In Table~\ref{distances} the geometric predictions of both versions of exchange and
the nonlocal correlation functional are compared with a most recent large
basis set MP2 calculation\cite{sinnokrot1} and with experiment.\cite{arunan}
The agreement shown by the HF version, vdW-DF(HF), is clearly outstanding.
For reference, the predictions of an earlier CCSD(T) calculation are also shown.\cite{hobza3}
A comparison of the binding energies is given in Table~\ref{energies}.

\newcommand{\supe}[1]{^{\mathrm{#1}}}		
\newcommand{\spa}[1]{^{\phantom{\mathrm{#1}}}}	

\begin{table}[t]
\vspace*{2em}
\begin{center}
\begin{tabular}{lccccc}
\hline
               	&CCSD(T)   &  vdW-DF 			&vdW-DF(HF)   			&   MP2  \\
\hline
Sandwich       	&  1.81    &  $2.37$			&  $2.77$     			&  3.64  \\
Slip parallel  	&  2.78    &  $\spa{a}2.80\supe{a}$	&  $\spa{b}4.48\supe{b}$    	&  4.95  \\
T-shape      	&  2.74    &  $2.28$			&  $4.38$     			&  3.63  \\
\hline\\
\end{tabular}
\caption{Binding energies in kcal/mol of benzene dimers in different configurations. 
Except where indicated the slip parallel configurations are slip-parallel-2.
The MP2 and CCSD(T) numbers are estimates given in Ref.~\onlinecite{sinnokrot1}
of the values that would be obtained in the basis set limit: these differ somewhat from
the depths of the MP2 and CCSD(T) curves in our figures [Model II of Ref.~\onlinecite{tsuzuki1}].
  $\supe{a}\,$The number for  the slip-parallel-1 
configuration is 2.74. $\supe{b}\,$The value for the slip-\hbox{parallel-1}
 configuration is
0.005 kcal/mole smaller.
\label{energies}
}
\end{center}
\end{table}

Thus, 
the DFT nonlocal correlation is a very efficient way to obtain MP2 level 
accuracy at a fraction of the computational cost with slightly varying 
results depending on the method for calculating all non-correlation energy.

\begin{figure*}[t]
\vspace*{1em}
\begin{center}{\includegraphics[width=.96\textwidth]{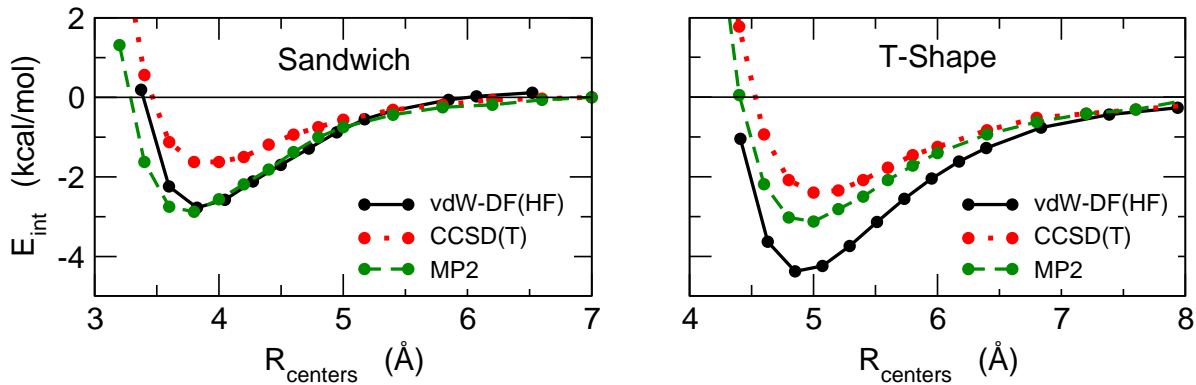}}\end{center}
\caption{Calculated interaction energies using the vdW-DF(HF) functional
[Eq.~(\ref{vdwdf}), with Hartree-Fock exchange substituted for
revPBE exchange].  The description of the plots
 is analogous to that for Fig.~2.
 }
\label{fig5}
\end{figure*}

\begin{figure}[htb]
\vspace*{1em}
\begin{center}{\includegraphics[width=0.48\textwidth]{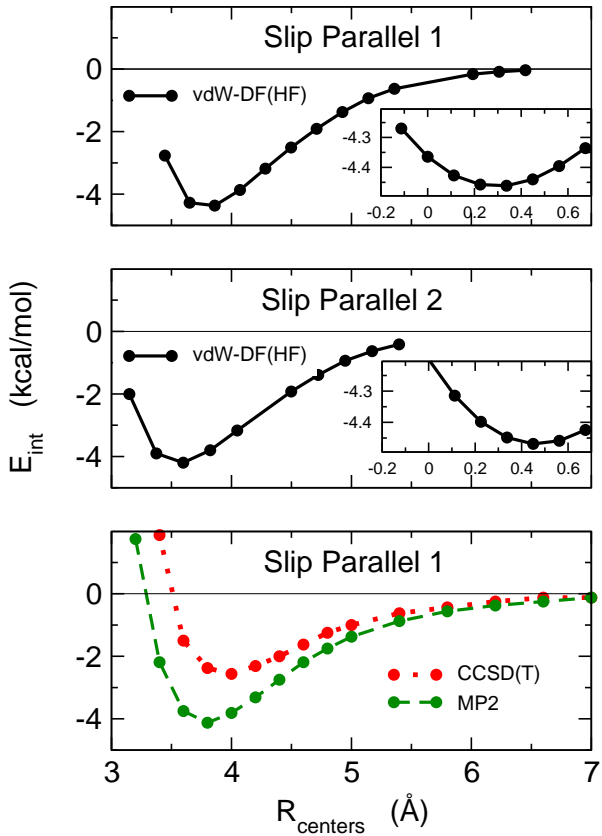}}\end{center}
\caption{Calculated interaction energies using the vdW-DF(HF) functional.
 The description of the plots
 is analogous to that for Fig.~3.}
\label{fig6}
\end{figure}

\section{Conclusions}
In conclusion, we have applied a nonlocal correlation functional capable of describing 
dispersion interactions to the benzene dimer.  We find that slip parallel is the lowest 
energy structure although all three geometries are nearly isoenergetic.  Our binding 
energies and qualitative predictions are consistent with CCSD(T) and MP2 calculations, 
although, our equilibrium geometries are slightly larger; however, the discrepancies in 
geometry are due to differences between revPBE and HF, not the nonlocal correlation 
energy functional.  Our method thus gives useful values for spacings and energies,
at a tiny fraction of the cost of the wave function methods. Specifically this cost
is comparable to that for calculations of the GGA type,
which typically give sporadic and often unphysical results for systems whose properties 
depend on the long range van der Waals interaction.

This benchmarking study has provided the evidence that this functional may 
be used in and even larger systems.  The consistency of our results 
with state-of-the-art wavefunction calculations allows us to pursue much 
larger and sometimes infinite systems.  Indeed, the functional has already 
been applied by the Chalmers part of our collaboration/cooperation to a 
physisorbed molecule on an infinite surface,\cite{benz-graph} to an 
infinite polyethylene crystal,\cite{pe} and to naphthalene, anthracene, 
and pyrene dimers\cite{svetla2}.  The two infinite system examples are cases 
in which other functionals that have been applied fail completely and in which 
the experimental data is sparse (one or two data points) with error bars that 
are larger than optimal.  The dimer examples were compared with CCSD(T) 
calculations for naphthalene\cite{tsuzuki-naphthalene} with favorable results.  
As a result,  we plan to continue with comparisons to other wavefunction 
calculations when possible, even when such calculations are not completely 
converged, and to move to even much larger systems such as base pairs of DNA.

\begin{acknowledgments}
The authors thank Svetla Chakarova-K\"ack,
Jesper Kleis, Elsebeth Schr{\"o}\-der, Per Hyld\-gaard, and
Bengt Lund\-qvist for helpful communications 
on ben\-zene and cal\-cu\-lat\-ions with the non\-local func\-tion\-al
and Ti\-mo Thon\-hauser for helpful computational programs to 
speed up the work.  This work was supported in part by NSF Grants DMR-0093079
and DMR-0456937.
\end{acknowledgments}


\begin{thebibliography}{37}
\bibitem[{opt()}]{optimize}
\bibinfo{note}{The currently used code for evaluating
  $E_\mathrm{c}^\mathrm{NL}$ is not optimized and uses a uniform grid, which is
  wasteful in the large low density regions. We believe that a substantial
  speedup is possible.}
\end{thebibliography}


\end{document}